\documentclass[prb,reprint,twocolumn,aps,amssymb,graphicx,floatfix,amsmath,superscriptaddress,showpacs,longbibliography]{revtex4-1}
\usepackage{bm}
\usepackage{float}
\usepackage{graphicx}
\usepackage{color}
\usepackage{amsfonts}
\usepackage{textcomp}
\usepackage{subcaption}
\usepackage{hyperref}
\usepackage{comment}
\hypersetup{colorlinks=true,linkcolor=blue,anchorcolor=blue,citecolor=blue,filecolor=blue,urlcolor=blue,bookmarksnumbered=true,pdfview=FitB}
\newcommand{\be}{\begin{equation}}
\newcommand{\ee}{\end{equation}}
\newcommand{\beq}{\begin{equation}}
\newcommand{\eeq}{\end{equation}}
\newcommand{\bea}{\begin{eqnarray}}
\newcommand{\eea}{\end{eqnarray}}
\newcommand{\bse}{\begin{subequations}}
\newcommand{\ese}{\end{subequations}}

\begin{document}

\title{
 The pairing glue in
 cuprate superconductors from the self-energy revealed via machine learning}

\author{Andrey V. Chubukov}

\affiliation{School of Physics and Astronomy and William I. Fine Theoretical Physics
Institute, University of Minnesota, Minneapolis, Minnesota 55455,
USA}

\author{J\"org Schmalian}

\affiliation{Institute for Theory of Condensed Matter, Karlsruhe Institute
of Technology, 76131 Karlsruhe, Germany}

\affiliation{Institute for Quantum Materials and Technologies, Karlsruhe Institute
of Technology, 76021 Karlsruhe, Germany}
\begin{abstract}
 Recently,  machine learning was applied to extract both  the normal and the anomalous
  components of the  self-energy
   from  photoemission data at the antinodal points in
   ${\rm Bi-}$based
   cuprate high-temperature superconductors [Y. Yamaji {\it et al.}, arXiv:1903.08060].
 It was argued that
  both components do show prominent peaks near
  $50\,{\rm meV}$,
   which hold information about the
   pairing glue,
  but the peaks  are hidden in the actual data, which measure only the
  total self-energy.
    We analyze the self-energy within an effective fermion-boson theory.
     We show that soft thermal fluctuations give rise to peaks in both  components of the self-energy at a frequency comparable to superconducting gap, while
     they cancel in the total self-energy; all irrespective of the nature of the pairing boson.  However,  in the quantum limit $T \to 0$ prominent peaks survive only
  for a very restricted subclass of pairing interactions.
  We argue that the way to potentially  nail down the pairing boson is to determine the thermal evolution of the peaks.
\end{abstract}
\maketitle

{\em Introduction:} The analysis of fine structures in the single-particle excitation
spectrum
 is one of  most reliable approaches
  to identify the interaction responsible for the pairing in superconductors. The prime example
is the  identification of  phonons
 as  pairing bosons
  in lead via their
 fingerprints
  in the single-particle tunneling density of states (TDOS)~\cite{Scalapino1966,Scalapino1969,McMillan1969}.
There have been several attempts to extend this logic to cuprate
superconductors, analyzing tunneling, angle-resolved photoemission (ARPES), optical, or inelastic neutron scattering spectra~\cite{ZXSchrieffer1997,Carbotte1999,Abanov2001,Vekhter2003,Dahm2009,Heumen2009,Choi2011,Heumen2012,Bok2016}.
  Such approaches are particularly challenging for the
    cuprates because i) the d-wave pairing interaction  depends on both momentum and frequency and  ii) the pairing interaction  likely changes due to feedback from superconductivity.

\begin{figure*}
\centering
\includegraphics[scale=0.8]{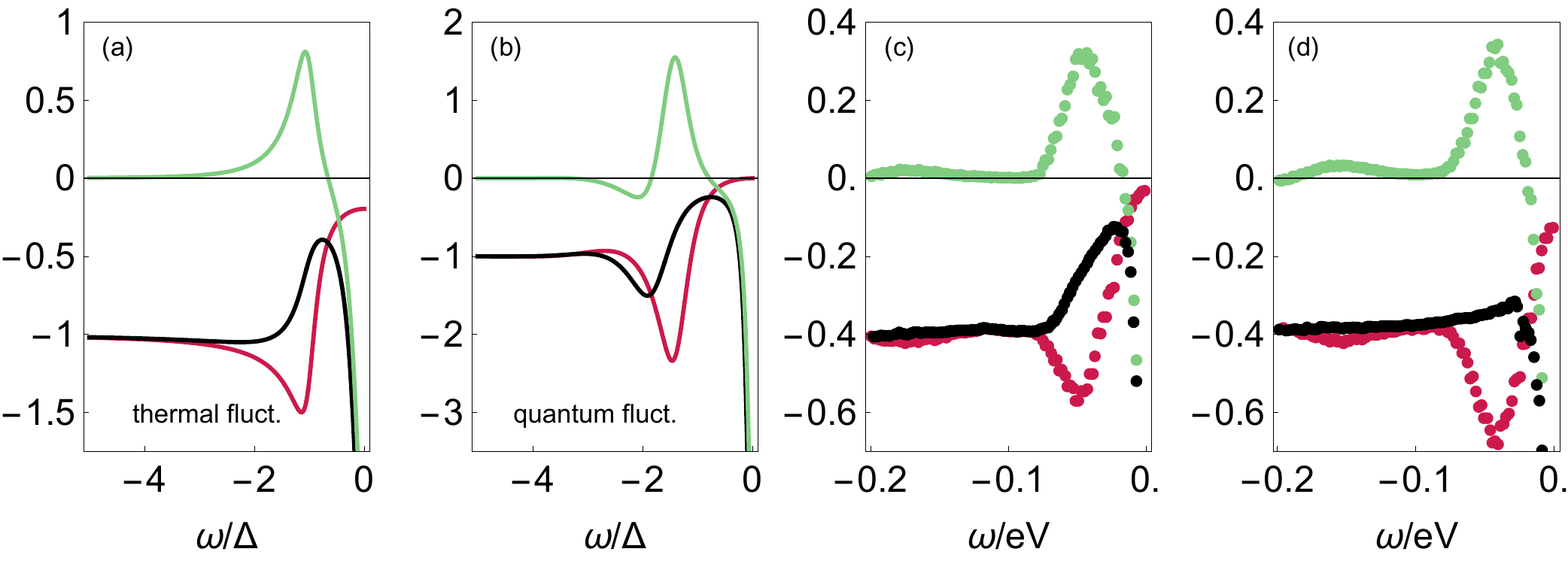}
\caption{Imaginary parts of the normal (red), total (black), and pairing (green) self-energies $\Sigma(\omega)$,  $\Sigma_{\rm tot}(\omega)$, and $W(\omega)$, respectively, obtained  for a pairing interaction with thermal fluctuations (a), quantum fluctuations and $\gamma=2$  (b) and from ARPES experiments using the machine-learning  analysis  of Ref.\onlinecite{Yamaji2019} for  Bi2212 (c) and Bi2201 (d).}
\label{Fig1}
\end{figure*}

In a superconducting state two distinct self-energies emerge due to
  mixing of electrons and holes\cite{Gorkov1958,Anderson1958,Nambu1960,Eliashberg1960}. This  gives rise to a matrix $2 \times 2$ structure of the fermionic Green's function. To determine the pairing
   glue
   it is highly desirable to have information about both, the normal  self-energy  $\Sigma_{\bf k}\left(\omega\right)$
and  the anomalous self-energy $\Phi_{\bf k}\left(\omega\right)$.
In a situation when the exchange of the same boson accounts for the pairing and for fermionic incoherence,
 the two
 self-energies are related.
  If both are extracted
 from the data, one can either verify - with a higher precision - the applicability of a given
  boson-mediated interaction, or put   strong constraints on the pairing glue
   without assuming a particular scenario.
 However, single-particle probes, like TDOS and ARPES,  provide information solely about
  the normal component of the Green's function $G_{\bf k} (\omega)$, which depends only on  the combination
$\Sigma_{\bf k}^{\left({\rm tot}\right)}\left(\omega\right)=\Sigma_{\bf k}\left(\omega\right)+W_{\bf k}\left(\omega\right)$,
where
\begin{equation}
W_{\bf k}\left(\omega\right)=\frac{\Phi_{\bf k}^{2}\left(\omega\right)}{\omega+
\epsilon_{\bf k}+\Sigma_{\bf k}^{*}\left(-\omega\right)}
\label{ch_1}
\end{equation}
contains the information about $\Phi_{\bf k}\left(\omega\right)$. To deduce two complex functions $\Sigma$ and $\Phi$ from only one observable seems hopeless
at first glance.
Recently, however, Yamaji \emph{et al.}\cite{Yamaji2019} argued that
 this can be achieved using a Boltzmann machine-learning approach.  They analyzed
 ARPES spectra~\cite{Kondo2009,Kondo2011} for optimally doped
 Bi2212
 with $T_{c}\approx90\:{\rm K}$ and  underdoped
Bi2201
with $T_{c}\approx29\:{\rm K}$, both  at  antinodal momenta ${\bf k}_{\rm a.n.} \approx (0,\pi)$. In both cases they found
 sharp structures
 in $\Sigma_{{\bf k}_{\rm a.n.} }\left(\omega\right)$
and $\Phi_{{\bf k}_{\rm a.n.} }\left(\omega\right)$
 at a frequency near $50\:{\rm meV}$, which is close to the value of the antinodal gap $\Delta_{\rm a.n.}$
(panels (c) and (d) of Fig.\ref{Fig1}). They further argued that
 these structures
  do not appear
   in the total self-energy
 $\Sigma_{{\bf k}_{\rm a.n.} }^{\left({\rm tot}\right)}\left(\omega\right)$
 (black data in the panels).

In this communication we compare the results of Ref.\onlinecite{Yamaji2019} with the forms
 of $\Sigma_{{\bf k}_{\rm a.n.} }\left(\omega\right)$
and $\Phi_{{\bf k}_{\rm a.n.} }\left(\omega\right)$ obtained in  various quantum-critical models of $d$-wave pairing due to  a dynamic interaction $V_{\bf q} (\Omega)$,
  mediated by a soft boson.
   We show that thermal fluctuations give rise to peaks in both  $\Sigma_{{\bf k}_{\rm a.n.} }\left(\omega\right)$
and $\Phi_{{\bf k}_{\rm a.n.} }\left(\omega\right)$
at a frequency slightly above $\Delta_{\rm a.n.}$,
     and these peaks do cancel in the total self-energy.  The peaks are present for any form of $V_{\bf q} (\Omega)$, as long as a boson is soft and $V_{\bf q} (0)$ is large,
      and in this respect do not
      place constraints on a
       pairing interaction.
  They do,
    nevertheless,  strongly support the view  that in optimally doped and underdoped cuprates the pairing boson is
    soft.
     Our results for the self-energies due to thermal fluctuations  are shown in Fig.\ref{Fig1} (a).
We can draw much stronger conclusions if
 singular structures in the self-energies
  survive in the quantum limit $T \to 0$.  In this
    case it must hold that the momentum-averaged pairing interaction
\begin{equation}
{\bar V} (\Omega) = N_F \oint V_{\bf q} (\Omega)\sim 1/|\Omega|^\gamma
\label{Eq_VOmega}
\end{equation}
is governed by an exponent $\gamma >1$.
  This
  would exclude  Landau-overdamped spin or charge fluctuations with Ornstein-Zernike form of the static $V_{\bf q} $,
 or pairing due to nematic fluctuations,
  as for these theories $\gamma \leq 1$, even when we include the feedback from superconductivity on a pairing boson.
       On the other hand this condition is satisfied for the pairing by nearly dispersionless boson,
        in which case $\gamma =2$ and
$ V_{\bf q} (\Omega) = V ({\bf q})/(\Omega_{\rm bos}^2 - \Omega^2)$,
  where $\Omega_{\rm bos}$ is small and $V ({\bf q})$ has an attractive  d-wave component. This is the case for pairing by
   a
       soft optical phonon\cite{Marsiglio1991,Karakozov1991,Combescot1995} and the strong coupling limit
        in a model of dispersionless fermions randomly coupled to optical phonons
          \cite{Esterlis2019,Wang2020}.
In Fig.\ref{Fig1} (b) we show the self-energies for $\gamma=2$ in the quantum regime.

The
frequency dependence of all self-energies in panels (a) and (b)
is in
 good
 agreement with the behavior obtained via machine learning,  panels (c) and (d).
 The self-energies
 in Bi2212 and Bi2201,  extracted in Ref.\onlinecite{Yamaji2019}, were obtained at $T=11\:K$ and $T=12\:K$, respectively\cite{Kondo2009,Kondo2011}, which
   are significantly below $T_{c}$.  This indicates that the  sharp features may
    be due to quantum
    fluctuations and thus do impose the restriction on the pairing mechanism.

Our scenario differs from the one presented in Refs. \cite{Imada_1,Imada_2}. In their approach, peaks in
$\Sigma_{{\bf k}_{\rm a.n.} }\left(\omega\right)$
and $\Phi_{{\bf k}_{\rm a.n.} }\left(\omega\right)$ originate from Mott physics, and the pole in $\Sigma_{{\bf k}_{\rm a.n.} }\left(\omega\right)$ exists already in the normal state.
In our scenario, the peak position is associated with the pairing gap and exists in the superconducting and pseudogap states, as long as pseudogap can be viewed as precursor to superconductivity.

{\em The model:} We use the Nambu-Gor'kov formalism with spinor $\psi\left(x\right)=\left(\psi_{\uparrow}\left(x\right),\psi_{\downarrow}^{\dagger}\left(x\right)\right)^{T}$,
 where
 $x=\left(\mathbf{x},t\right)$ combines coordinates and time.
The Green's function ${\cal G}\left(x,x'\right)=-i\theta\left(t-t'\right)\left\langle \left[\psi\left(x\right),\psi^{\dagger}\left(x'\right)\right]\right\rangle $
is a $\left(2\times2\right)$ matrix.  Fourier transformation to the momentum
and frequency representation yields
\begin{equation}
{\cal G}_{\bf k}^{-1}\left(\omega\right)=\left(\begin{array}{cc}
\omega-\epsilon_{\bf k}-\Sigma_{\bf k}\left(\omega\right) & \Phi_{\bf k}\left(\omega\right)\\
\Phi_{\bf k}\left(\omega\right) & \omega+\epsilon_{\bf k}+\Sigma_{\bf k}^{*}\left(-\omega\right)
\end{array}\right),
\end{equation}
with bare dispersion $\epsilon_{\bf k}$ and two
self-energies, $\Sigma_{\bf k}\left(\omega\right)$
and $\Phi_{k}\left(\omega\right)$. The
total self-energy $\Sigma_{\bf k}^{\left({\rm tot}\right)}$  is defined via $G^{-1}_{\bf k} \left(\omega\right) = \omega - \epsilon_{\bf k} - \Sigma_{\bf k}^{\left({\rm tot}\right)}\left(\omega\right)$,
where $G_{\bf k} \left(\omega\right)$ is the  upper-left element of ${\cal G}_{\bf k} \left(\omega\right)$ and $\Sigma_{\bf k}^{\left({\rm tot}\right)}\left(\omega\right)$,
 is given by Eq. (\ref{ch_1}).
 In what follows we assume particle-hole symmetry $\Sigma_{\bf k}^{*}\left(-\omega\right) = - \Sigma_{\bf k}\left(\omega\right)$. This assumption
 breaks down
  near an antinodal point at energies above a few hundred meV,  but holds
   at energies $\omega
   \sim \Delta_{\rm a.n.}$.

 To obtain the coupled equations for $\Sigma_{\bf k}\left(\omega\right)$ and $\Phi_{\bf k}\left(\omega\right)$
  in closed form we further assume, following earlier works\cite{Abanov2001},
   that corrections to side vertices in the diagrams for $\Sigma$ and $\Phi$ can be neglected. This allows one to
    obtain a set of self-consistent equations for
    the two self-energies.
  The coupled equations are expressed most naturally in the Matsubara formalism, where we have
     \begin{eqnarray}
\Sigma_{\bf k}\left(i\omega_m\right) & = & -\int_{{ k}'}V_{{\bf k}-{\bf  k}'}\left(i\omega_m-i\omega_{m'}\right)\frac{i\omega_{m'}-\Sigma_{{ \bf k}'}\left(i\omega_{m'}\right)}{{\cal N}_{{\bf  k}'}\left(i\omega_{m'}\right)},\nonumber \\
\Phi_{\bf k}\left(i\omega_m\right) & = & \pm\int_{{ k}'}V_{{\bf k}-{\bf k}'} \left(i\omega_m-i\omega_{m'}\right)\frac{\Phi_{{\bf k}'}\left(i\omega_{m'}\right)}{{\cal N}_{{\bf k}'}\left(i\omega_{m'}\right)}.
\label{eq:Eliashberg full}
\end{eqnarray}
We defined
 ${\cal N}_{\bf k}\left(i\omega_m\right)=\epsilon_{\bf k}^{2}+\left(\omega_m+i\Sigma_{\bf k}\left(i\omega_m\right)\right)^{2}+\Phi_{\bf k}\left(i\omega_m\right)^{2}$
and $\int_{k}\cdots=T\sum_{\omega_m}\int\frac{d^{2}k}{\left(2\pi\right)^{2}}\cdots$. The upper (the lower) sign refers to the charge (spin)
channel. Finally, we assume that the pairing interaction is  attractive in the $d$-wave
channel and that self-energies for antinodal fermions at ${\bf k} = { \bf k}_{\rm a.n.}$, $\Sigma_{{ \bf k}_{\rm a.n.}}\left(i\omega_m\right) = \Sigma\left(i\omega_m\right)$ and
$\Phi_{{ \bf k}_{\rm a.n.}}\left(i\omega_m\right) = \Phi\left(i\omega_m\right)$, are
 primarily determined by internal fermions with
  momenta ${\bf k}'$ that are also near one of
   antinodal points.

 At this stage we are agnostic about the nature of the pairing interaction. It  might come from spin or charge fluctuations (or the combination of the two)\cite{Abanov2001,Bickers1989,Monthoux1993,Grilli1991,Perali1996},
 from nematic fluctuations\cite{Husemann2012,Yamase2013,Maier2014,Lederer2015,Metitski2015,Kang2016},  from fluctuations of a current order parameter\cite{Varma1997}, or  from soft phonons\cite{Johnston2012,Esterlis2018,Esterlis2019,Wang2020}. The exponent $\gamma$ of Eq.\ref{Eq_VOmega} depends on the mechanism under consideration; see above.
In all cases, $V_{\bf q} (\Omega)$ gives rise to a strong attraction in the pairing channel,
 and, simultaneously, to a large
    fermionic self-energy $\Sigma_{{\bf k}}\left(\omega\right)$, which makes fermions incoherent.  The incoherence and the pairing originate from the same $V_{\bf q} (\Omega)$ and compete with each other.    The competition gives rise to structures in both, $\Sigma_{{\bf k}}
    \left(\omega\right)$
and $\Phi_{{\bf k}
}\left(\omega\right)$  that must be simultaneously understood.

 We consider two complementary limits with regards
to the electronic dispersion $\epsilon_{\bf k}$. One is the case of a strong dispersion, where the momentum integration can be performed similar to the usual
 Eliashberg theory. In the opposite limit the dispersion near the antinodes
is so flat that it can be ignored all-together (Ref. \onlinecite{Esterlis2019}).  This second case has been extensively studied recently in the context of Sachdev-Ye-Kitaev-type models\cite{Esterlis2019,Wang2020}.  Interestingly we find quantitatively the same behavior in both limits.
Below we first present the results for a strong and then for a flat dispersion.

 {\em Thermal fluctuations:}  The thermal contributions to $\Sigma$ and $\Phi$ come from the terms with $\omega_m = \omega_{m'}$ in (\ref{eq:Eliashberg full}).
These terms contain the static interaction  $V_{{\bf k}-{\bf k}'} (0)$, which is
large when the pairing boson is soft (e.g., $V_{\bf q} (0) = V ({\bf q})/\Omega_{\rm bos}^2$  for  $\gamma=2$).
%in Eq.\,\ref{gamma2}).
Introducing, as usual, $\Sigma (i\omega_m) = i\omega_m (1 - Z(i\omega_m))$ and
$\Phi (i\omega_m) = Z(i \omega_m) \Delta (i \omega_m)$ and singling out the thermal piece, we obtain from  (\ref{eq:Eliashberg full})
\begin{widetext}
\bea
&&Z (i\omega_m) \approx  \pi T \frac{{\bar V} (0)}{\sqrt{(\omega_m)^2 + (\Delta (i\omega_m))^2}} , \label{ch_2a} \nonumber \\
&&\Delta (i \omega_m) = \pi T \sum_{\omega_m'}\frac{{\bar V} (i\omega_m - i \omega_m')}{\sqrt{(\omega_m')^2 + (\Delta (i\omega_m'))^2}} \left(\Delta (i\omega_m') - \Delta (i \omega_m) \frac{\omega_m'}{\omega_m} \right).
\label{ch_2}
\eea
\end{widetext}
\begin{figure*}
\centering
\includegraphics[width=0.45\hsize]{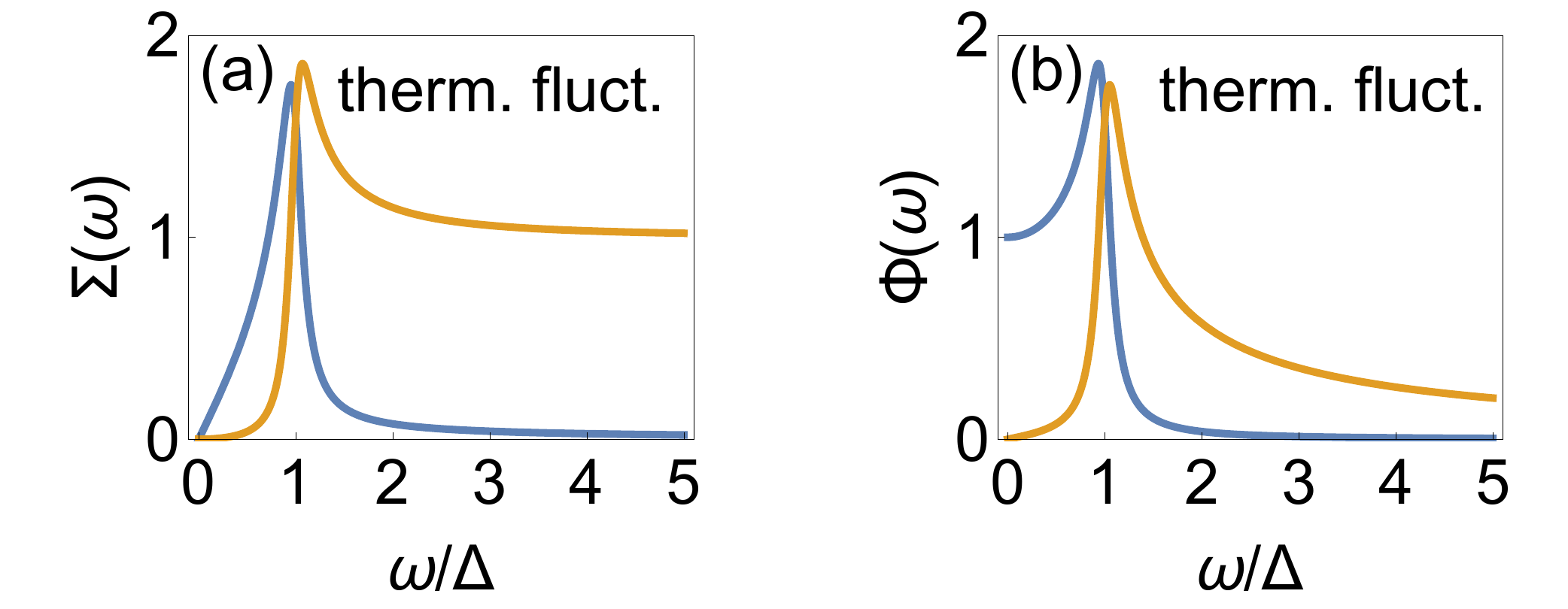}\includegraphics[width=0.45\hsize]{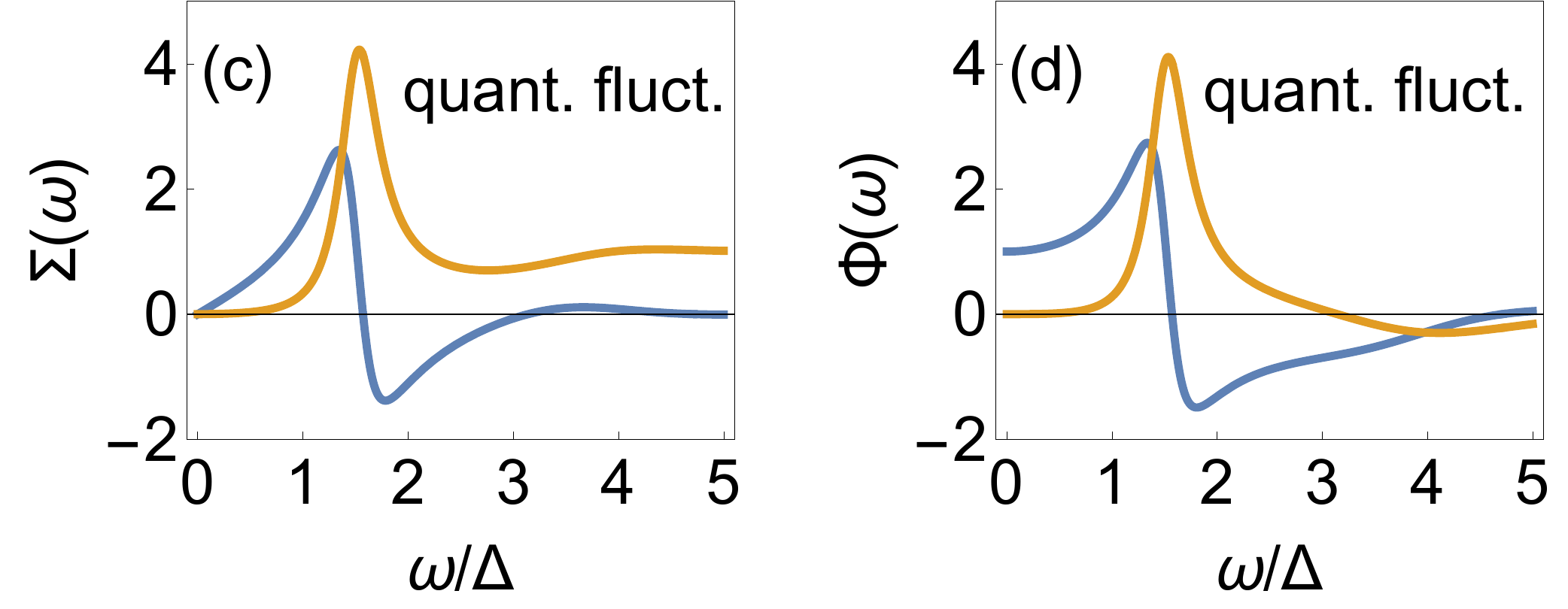}
\caption{The real (blue) and imaginary (yellow) parts of the normal and anomalous self-energy, $\Sigma (\omega)$
(panels a and c)
and $\Phi (\omega)$ (panels b and d). Panels a,b --  at a finite temperature, when the dominant contributions to both self-energies come from thermal fluctuations.
Panels c,d -- at $T=0$ for the case  $\gamma =2 $ (see text).
The frequency is in units of $\Delta$ and
 fermionic damping is
  $0.1 \Delta$.
The peaks in Im $\Sigma$ and Im $\Phi$ at $\omega \leq \Delta$ are clearly visible.
 At higher frequencies, Im $\Sigma (\omega)$ saturates, other self-energies drop, and, in the  $T=0$ case,  ${\rm Re} \Sigma (\omega)$ and ${\rm Re} \Phi (\omega)$ change sign.}
\label{fig:A}
\end{figure*}

We see from Eq.(\ref{ch_2}) that the thermal contribution ${\bar V} (0)$ determines the value of $Z (i \omega_m)$  but cancels out in the r.h.s. of the equation for $\Delta (i\omega_m)$.  As the result, $\Delta (i \omega_m)$ is determined by non-thermal terms and remains non-singular even when ${\bar V} (0)$ diverges.
 The cancellation holds for the same  reason as the Anderson theorem,  because thermal fluctuations scatter with zero frequency transfer and a finite momentum transfer act as non-magnetic impurities.
Because of this cancellation,
$\Delta (i \omega_m)$ remains non-singular even at criticality and at small  $|\omega_m|$
can be approximated
by  a frequency-independent constant $\Delta$. This approximation holds for  $|\omega_m| \leq \Delta$.
Performing the analytical continuation to the real axis
 we then obtain
 normal and anomalous self-energies in the form
\bea
 \Phi (\omega) / \Delta&=&-\Sigma (\omega)/\omega \approx  \pi T {\bar V}(0) \frac{1}{\sqrt{\Delta^2- (\omega + i \delta)^2}} ,\nonumber \\
  \Sigma^{\left({\rm tot}\right)}\left(\omega\right) &\approx& \pi T {\bar V}(0) \frac{\omega}{\sqrt{\Delta^2 - (\omega + i \delta)^2}} \left(1 - \frac{\Delta^2}{\omega^2}\right).
 \label{ch_3a}
 \eea
 We see that
  both $\Sigma (\omega)$ and $\Phi (\omega)$ in Eq.(\ref{ch_3a}) are singular at $\omega \approx \Delta$, but the singularities cancel out in the total self-energy because of the factor $\omega^2 - \Delta^2$.
More detailed calculations, in which sub-leading terms in $Z(i \omega_m)$ are kept~\cite{Wu2019}, give a very similar result, only the damping $\delta$ becomes finite
and the singularity shifts to somewhat larger $\omega$.
We plot the real and imaginary parts of $\Sigma (\omega)$ and $\Phi (\omega)$ in panels a,b in Fig. (\ref{fig:A}) and in
panel (a) of Fig.\ref{Fig1}. We see that ${\rm Im} \Sigma (\omega)$ and Im $\Phi (\omega)$ have peaks at $\omega$ slightly above $\Delta$. At larger $\omega$, ${\rm Im} \Sigma (\omega)$ saturates, while ${\rm Im} \Phi (\omega)$
 rapidly drops.  ${\rm Re} \Sigma (\omega)$ is linear in $\omega$ at small frequencies and ${\rm Re} \Phi (\omega)$ is finite. Both pass through maximum at
 $\omega
  \approx  \Delta$ and
  then rapidly drop and are already small at a frequency where ${\rm Im} \Sigma (\omega)$ has a maximum. All these features are also present in the normal and anomalous self-energies extracted in Ref.\onlinecite{Yamaji2019}.

 We emphasize that the existence of peaks in ${\rm Im} \Sigma (\omega)$ and ${\rm Im} \Phi (\omega)$
 due to thermal fluctuations,
 and the cancellation of the peak in $\Sigma^{\left({\rm tot}\right)}\left(\omega\right)$ are generic properties of any model of pairing by a soft boson.
   They are present as long the temperature is not too small such that ${\bar V} (0) \gg {\bar V} (2\pi T)$.

 {\em Quantum fluctuations:}  At $T=0$ thermal fluctuations are absent, and one has to integrate over frequency in (\ref{eq:Eliashberg full}) instead of summing up over a discrete set of $\omega_{m'}$.
To check whether the features in the self-energies are still present, we
focus on the quantum-critical point, where these features are expected to be the strongest.
In this case,  ${\bar V} (\Omega_m) = g/|\Omega_m|^\gamma$, and
\bea
Z (i\omega_m) &=&1 +  \frac{g}{\omega_m} \int \frac{d \omega'_m}{|\omega_m-\omega'_m|^\gamma} \frac{\omega'_m}{\sqrt{(\omega'_m)^2 + (\Delta (i\omega'_m))^2}} ,
\nonumber \\
\Delta (i \omega_m) &=& g \int \frac{d \omega'_m}{|\omega_m-\omega'_m|^\gamma} \frac{\Delta (i\omega'_m) - \Delta (i \omega_m) \frac{\omega'_m}{\omega_m}}
{\sqrt{(\omega'_m)^2 + (\Delta (i\omega'_m))^2}} .
\label{ch_6}
\eea
The gap
 function still saturates at
 $\Delta (i \omega_m) \approx \Delta$ at $|\omega_m| \leq \Delta$. Approximating $\Delta (i\omega_m)$ by $\Delta$ in the r.h.s. of equation for $Z(i \omega_m)$,
  evaluating the integral, and converting the result to real frequencies, we find that $Z(\omega)$ is featureless
   in all models with
    $\gamma <1$.

The situation changes for $\gamma >1$. Now the frequency integral in (\ref{ch_6}) diverges. One can cut-off the divergency by
 moving a  system slightly away from the critical point, where
 $V^* = \int d \Omega {\bar V} (\Omega)$
is
finite. We then obtain
\beq
Z (i\omega_m) \approx    \frac{V^*}{\sqrt{(\omega_m)^2 + (\Delta (i\omega_m))^2}}.
\label{ch_7}
\eeq
 Comparing Eqs.(\ref{ch_7}) and (\ref{ch_2a}),
  we see that the functional form of $Z(i\omega_m)$ is the same in both cases. As a consequence, for $\gamma >1$, the normal and anomalous self-energies
   in real frequencies show the same features at $T=0$ due to quantum fluctuations as at $T >0$ due to classical, thermal fluctuations.

The analysis
 at $T=0$ can be further advanced for $\gamma =2$
In this case the solution for $\Delta (\omega)$ in real frequencies is highly unconventional~\cite{Carbotte1990,Marsiglio1991,Karakozov1991,Combescot1995}. Namely,
  it  does saturate at
  a finite $\Delta$ at frequencies $\omega \leq \Delta$, but at larger $\omega$ behaves as
 $\Delta (\omega) \propto \Delta \omega/\sin{\psi (\omega + i \delta)}$, where $\psi (\omega+ i \delta)$ is an increasing function of $\omega$,
  approximately linear in $\omega$
  (Ref.\,\onlinecite{Combescot1995}).
    In this situation,
  the self-energies behave as
 \bea
 \Sigma (\omega) &\approx& - V^* \tan{\psi (\omega + i \delta)} + ..., \nonumber \\
\Phi (\omega) &\approx & \frac{V^*}{\cos{\psi (\omega + i \delta)}} + ... ,\nonumber \\
 \Sigma^{\left({\rm tot}\right)}\left(\omega\right) &\approx &- V^* \cot{\psi (\omega + i \delta)} + ...\, .
\label{ch_8}
\eea
where dots stand for non-singular terms. For infinitesimally small $\delta$, ${\rm Im} \Sigma (\omega)$ and ${\rm Im} \Phi (\omega)$ have strong peaks at $\omega$  where $\psi (\omega) = \pi/2 + n\pi$. At these frequencies, the singular part of $\Sigma^{\left({\rm tot}\right)}\left(\omega\right)$ vanishes. Conversely, when $\psi (\omega) = m\pi, m \neq 0$, ${\rm Im} \Sigma^{\left({\rm tot}\right)}\left(\omega\right)$ has a peak, while the singular parts of $\Sigma (\omega)$ and $\Phi (\omega)$ vanish.
 For a finite $\delta$, the peaks get broadened and
  only the peak at $\omega \approx \Delta$ remains.      In  panels c,d in Fig. (\ref{fig:A})
  and in panel (b) of Fig.\ref{Fig1} we  show
 the self-energies using
 $\psi (\omega + i \delta) = \omega + i \delta$
 and
  $\delta =\delta  (\omega) =0.1\omega^2/\Delta$.
   We see that the behavior is qualitatively similar to that in Fig. \ref{fig:A}, the
     only difference is that ${\rm Re} \Sigma (\omega)$ and ${\rm Re} \Phi (\omega)$ change sign at $\omega \geq \Delta$.  Interestingly, ${\rm Re} \Sigma (\omega)$ and ${\rm Re} \Phi (\omega)$,  extracted by in Ref.\onlinecite{Yamaji2019} also
     change sign at $\omega \geq \Delta$.

  The same set of results is obtained if one completely ignores the electronic dispersion.
In this limit the equations for $Z (i \omega_m)$ and $\Delta (i \omega_m)$ become, at $T=0$,
\begin{widetext}
\bea
&&Z (i\omega_m)  = 1+ \frac{g}{\omega_m} \int \frac{d \omega_m'}{|\omega_m-\omega'_m|^\gamma} \frac{\omega'_m}{Z (i \omega'_m) ((\omega'_m)^2 + (\Delta (i\omega_m))^2)} \nonumber \\
&&\Delta (i \omega_m) = g \int \frac{d \omega_m'}{|\omega_m-\omega'_m|^\gamma} \frac{\Delta (i\omega'_m) - \Delta (i \omega_m) \frac{\omega'_m}{\omega_m}}{
{ Z(i \omega'_m) \left((\omega'_m)^2 + (\Delta (i\omega'_m))^2\right)}}
\eea
\end{widetext}
Solving these equations and  converting the result to real frequencies, we find that $Z(\omega)$ is rather featureless for $\gamma <1$, and $\Sigma (\omega)$ and $\Phi (\omega)$ do not display sharp features at $\omega \sim \Delta$.  For $\gamma >1$, the frequency integral for $Z(i \omega_m)$ again diverges, and
\beq
Z (i\omega_m) \approx    \frac{V^{*}}{Z(i \omega_m)} \frac{1}{(\omega_m)^2 + (\Delta (i\omega_m))^2}. \nonumber
\eeq
Solving for $Z(i\omega_m)$ we find that it has the same functional form as for strong dispersion:
$Z(i \omega_m) \approx (V^*)^{1/2} /\sqrt{\omega_m^2 + (\Delta (i\omega_m))^2}$.
 Substituting $Z(i \omega_m)$ into the gap equation we obtain
\beq
 \Delta (i \omega_m) = g_{\rm eff} \int \frac{d \omega_m'}{|\omega_m-\omega'_m|^\gamma} \frac{\Delta (i\omega'_m) - \Delta (i \omega_m) \frac{\omega'_m}{\omega_m}}{
\sqrt{(\omega'_m)^2 + (\Delta (i\omega'_m))^2}}, \nonumber
\eeq
where $g_{\rm eff} = g/(V^*)^{1/2}$.
This equation is exactly the same as the gap equation (\ref{ch_6}) for strong dispersion.  Hence, the functional forms of normal and anomalous self-energies are the same as in Fig. (\ref{fig:A}).

{\em Conclusions:} The simultaneous extraction of the normal and anomalous self-energy from photoemission data, enabled by recent machine-learning approaches\cite{Yamaji2019},
 allows for significantly deeper insight into the nature of the dynamic pair formation.
  We argue that it implies that pairing is
  mediated by a soft, near-critical bosonic mode.  Furthermore, if sharp peaks in both self-energies  survive down  to the lowest temperatures
   (i.e., they are due to quantum excitations), their presence alone  imposes  strong restrictions on the energy dependence of a soft pairing
   boson.  We call for  a systematic analysis of the peaks
     as function of temperature.

   This work was supported by the Office
of Basic Energy Sciences, U.S. Department of Energy, under award
 DE-SC0014402 (A.V.C.) and the Deutsche Forschungsgemeinschaft (DFG) through the  SCHM 1031/7-1 (J.S.).
 % We thank all colleagues with whom we discussed this work.

\end{document}